# Bio-heterojunction Effect on Fluorescence Origin and Efficiency Improvement of Firefly Chromophores

*Duanjun Cai*[*,†]*, Miguel A. L. Marques*[†,‡]*, Bruce F. Milne*[†]*, Fernando Nogueira*[†]

CFC, Departamento de Física, Universidade de Coimbra, 3004-516, Coimbra, Portugal, and Laboratoire de Physique de la Matière Condensée et Nanostructures, Université Lyon I, CNRS, UMR 5586, Domaine scientifique de la Doua, F-69622 Villeurbanne Cedex, France

---

[†] Universidade de Coimbra

[‡] Université Lyon I

**Abstract**

We propose the heterojunction effect in the analysis of the fluorescence mechanism of the firefly chromophore. Following this analysis, and with respect to the HOMO-LUMO gap alignment between the chromophore's functional fragments, three main heterojunction types (I, II, and I*) are identified. Time-dependent density-functional theory optical absorption calculations for the firefly chromophore show that the strongest excitation appears in the deprotonated anion state of the keto form. This can be explained by its high HOMO-LUMO overlap rate due to strong bio-heterojunction confinement. It is also found that the nitrogen atom in the thiazolyl rings, due to its larger electronegativity, plays a key role in the emission process, its importance growing when HOMO and LUMO overlap at its location. This principle is applied to enhance the chromophore's fluorescence efficiency and to guide the functionalization of molecular optoelectronic devices.

**TOC**

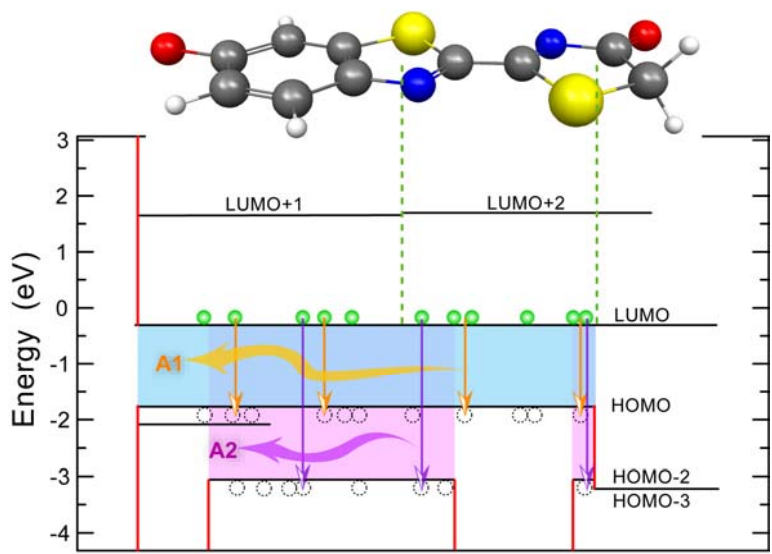

Fluorescent proteins have become a unique marking tool in molecular and analytical biology, attracting a large interest by the scientific community due to their wide present and potential applications in multiple disciplines. In particular, the well known fluorescent protein, firefly luciferase, has been successfully used in various research frontiers of bio-imaging,[1] as a reporter for ATP generation,[2] gene expression,[3] HIV dynamics,[4] motion of single-molecule motors,[5] and biosensors for environmental pollutants.[6] In the firefly bioluminescence process, the oxyluciferin (OxyLH2) molecule, as shown in Chart 1, has been doubtlessly identified as the important chromophore for the emission of visible light. Remarkably, the same molecule, OxyLH2 is also the crucial light emitter in other organisms, such as click beetles and railroad worms.[7-9] This suggests that "Nature knows" that OxyLH2 is a highly efficient light emitter, and moreover, that this occurs even in different microenvironments.

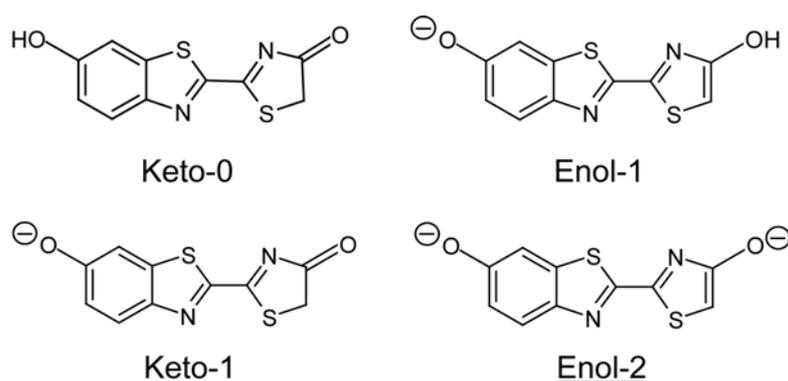

**Chart 1.** Chemical structures of important isomers (Keto- and Enol-form) of firefly chromophore, OxyLH2, in different protonated and deprotonated states.

There are several reasons that may explain the success of this chromophore. The OxyLH2 keto-form, 2-(6-hydroxy-1,3-benzothiazol-2-yl)-1,3-thiazol-4-one, consists of a hydroxybenzothiazolyl (HOBT) group and an oxythiazolyl (OT) group (see Chart 1). Thiazoles are aromatic and the aromaticity makes them relatively stable, both chemically and physically.[10] Also, as members of the heterocyclic family, thiazoles strongly allow for specific functionalizations. Furthermore, light emitted from OxyLH2 is in the visible region, making it a suitable candidate for visible fluorescence. Finally, the emission efficiency of OxyLH2 is high enough for long lifetime fluorescence at high brightness and for easier energy supply.

Clearly, understanding the mechanism of the OxyLH2 fluorescent process is important for better controlling and effectively improving light emission from this biological reporter. Therefore, it is not surprising that the elucidation of the mechanism of firefly bioluminescence has been the goal of many recent studies of the OxyLH2 molecules.[11-13] Biochemical studies have mainly focused on the microenvironment effects, such as the solvent effect (pH sensitivity, orientation polarizability, or hydrogen bond formation) and the residue interactions.[14-16] Although color shifts have been observed by changing the solvent[17] or by mutagenesis,[18, 19] a convincing mechanistic understanding of the process inside the molecule remains elusive. Another line of research consists in studying in detail the essential mechanism for light absorption and emission by focusing on gas-phase OxyLH2, and excluding influences from, e.g., enzyme or solvent.[20] Based on this idea, one can perform controlled experiments for the chromophores *in vacuo*, and compare them to accurate first-principles calculations. This path has been successfully used for, e.g., the green fluorescent protein.[21-23]

In this letter, we use the concept of a heterojunction to the study of the fluorescence of gas-phase OxyLH2. Even if the concept of a heterojunction is a fundamental tool in the understanding of, e.g., semiconductor structures, it is much less used in the field of biological chromophores. We think, however, that it is a very useful concept, bringing a completely different perspective to the understanding of the highly efficient light emission process in OxyLH2. Weak confinement effect on charges was observed due to the bio-heterojunctions formed by different fragments. This result highlights an important aspect of the fluorescence mechanism and explains the high emission efficiency of OxyLH2. A method for the improvement of the bioluminescence brightness is also proposed.

To build our model of a "bio-heterojunction" we use information extracted from time-dependent density functional theory (TDDFT)[24] calculations of the relevant chromophores. Previous studies[25, 26] of the mechanism for firefly bioluminescence have suggested that the final-step-product, OxyLH2, acts as the important light emitter of yellow-green light in the case of the North American firefly (*Photinus pyralis*).[27] It is widely believed that the OxyLH2 emitter is in its keto form, but chemiluminescence studies point to another important isomer, the enol form.[28] With the exception of the tautomeric changes at the keto/enol moiety, the two forms are electronically and structurally similar. We performed a systematic study of all the states of OxyLH2 possibly involved in firefly luminescence. They were labeled as Keto-0, Keto-1, Enol-1, and Enol-2 (Chart 1). Keto-0 is the protonated keto form, and Keto-1 is the deprotonated monoanion of OxyLH2, which is the one believed to be the emitter in bioluminescence.[29] Enol-1 is the phenolate ion of the enol form, whereas the dianion state, Enol-2, plays the key role for the light emission in chemiluminescence.[28] The reference structure of OxyLH2 molecule is taken from the X-ray structure of Japanese genji-botaru firefly luciferase (*Luciola cruciata*, PDB database code: 2d1r).[30]

The TDDFT absorption spectra of OxyLH2 in different states are shown in Figure 1. The first excitation energy (A1) of Keto-0 appears in the violet range, whereas that of Keto-1 is redshifted to the green color. This red shift by the deprotonation of the left-end hydroxyl group effectively moves the emitted light of OxyLH2 into the longer-wavelength visible band. In contrast, the A1 peak of Enol-1 is already in the green region.and its red shift in Enol-2 dianion brings it to the orange region. Although the wavelength of Enol-2 is closer to the real color seen in firefly bioluminescence, the energetic instability of the active site, mentioned by Nakatani and coworkers,[30] makes it unfavorable for the key emitting species in firefly fluorescent protein. Thus, we will start the discussion by focusing initially on the properties of keto forms alone.

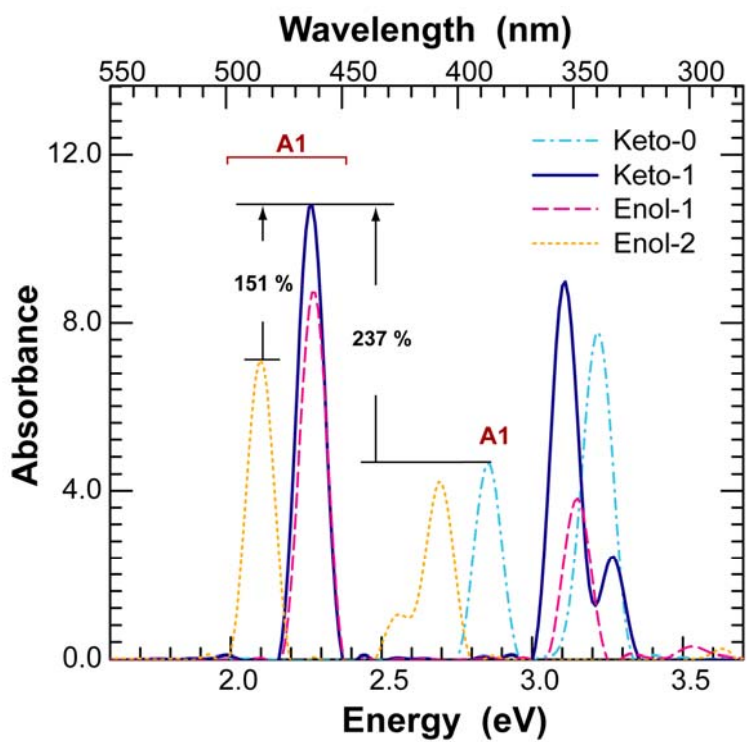

**Figure 1.** TDDFT absorption spectra of different isomer forms of OxyLH2. The intensity of the A1 peak of the Keto-1 is 237% higher than that of the keto-0 and 151% higher than that of the Enol-2.

**Table 1.** Contribution ratio $p$ and oscillator strength $f$ of HOMO→LUMO transitions for main spectral peaks, A1 and A2, in different OxyLH2 states.

| OxyLH2 | $p\ (f)$ | |
|---|---|---|
| | HOMO→LUMO (A1) | HOMO-2→LUMO (A2) |
| Keto-0 | 79% (0.161) | 62% (0.262) |
| Keto-1 | 78% (0.358) | 43% (0.329) |
| Enol-1 | 88% (0.219) | 86% (0.234) |
| Enol-2 | 94% (0.258) | 54% (0.116) |

A detailed analysis of the transition matrix elements obtained with Casida's method is summarized in Table 1, which shows the donor-acceptor molecular states involved in the transitions leading to the main spectral peaks and the corresponding transition probabilities. Peaks A1 and A2 are associated with transitions from the highest occupied molecular orbital to the lowest unoccupied molecular orbital (HOMO→LUMO) and HOMO-2→LUMO, respectively. The HOMO-1→LUMO transition matrix element is too small to give rise to an intense peak. Similarly, those states, such as LUMO+1 or higher, are not significantly involved in the transition for the A1 or A2 peaks. This can be explained in terms of the spatial distributions of the molecular orbitals involved in the transitions, which are listed and compared in the first column of Figure 2. The degree of spatial overlap of HOMO and LUMO orbitals plays a key role in the selection rules for these $\pi-\pi^*$ transitions. Larger HOMO-LUMO overlapping volumes (highlighted by rectangular boxes) lead to stronger peaks. The volume of the spatial overlap of different HOMO-LUMO orbitals can then be associated with the corresponding transition probability, and consequently with the allowed excitation energies of a molecule. The overlapping volume criterion can thus be viewed as a general selection rule for the molecular excitation spectra, in a very direct and simplistic description. Also, in the case of Keto-0, one may argue that the A1 transition takes place mainly at the HOMO-LUMO overlap on the C=N chain (Figure 2a). Similarly, one can claim that the A2 transition occurs at the HOMO-2→LUMO overlap at the nitrogen atomic site in the thiazolone ring on the right-hand side. These two critical transitions seem therefore to be highly associated with the nitrogen atoms of the heterocyclic rings. Similar conclusions can be drawn from the results for the enol forms.

The HOMO-LUMO overlapping volumes can certainly be different for different molecules or even different forms of the same molecule. From the comparison between the orbitals in Figure 2, it is clear that, for the Keto-1 form, the LUMO extends widely in both the left benzothiazolyl and the right thiazolyl rings, and the HOMOs are distributed across both rings as well. This may be due to a decrease of the OxyLH2 polarization along the long molecular axis, as a result of the deprotonation of the left hydroxyl end. From the distribution of LUMO and HOMO orbitals, it can be seen that the polarization in the Keto-0 molecule is aligned along the longest axis of the molecule from the benzothiazolyl to the thiazolyl fragment. The deprotonation in Keto-1 introduces a negative charge at the 6-hydroxyl group oxygen resulting in a decreased stabilization of the highest energy orbitals in this region of the molecule compared with the neutral Keto-0. Therefore, the π electrons in the benzothiazolyl ring are redistributed and extend further towards the thiazolone ring as is evidenced by the increased delocalization of the HOMO orbital. Meanwhile, the LUMO orbital in Keto-1 becomes more delocalized, spreading into both left and right moieties. The overlap between HOMO and LUMO in Keto-1 is significantly increased, an indication of this increase being the area of the rectangular boxes in Figure 2. The higher overlapping volume results in greater transition probability and therefore higher light emission efficiency and thus favors Keto-1 over Keto-0 as the light emitter in the high quantum-yield firefly bioluminescence reaction. The dominant overlapping regions are situated in the vicinity of the C2-C2' bridging bond between the thiazolyl rings, suggesting that this may be the most crucial region in determining the efficiency of oxyluciferin's fluorescence activity.

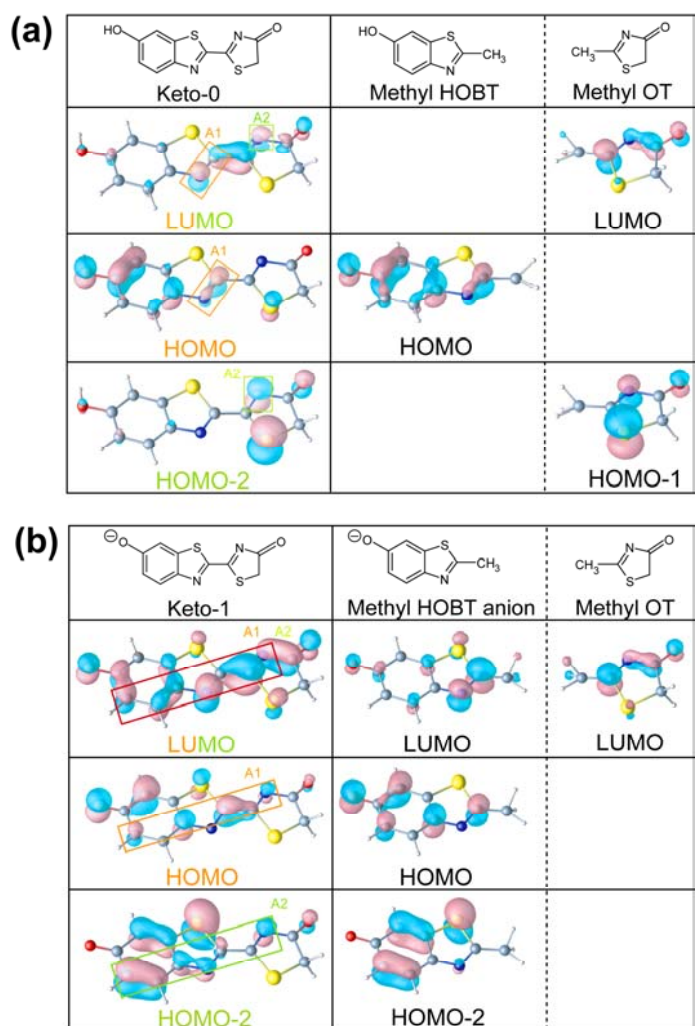

**Figure 2.** Comparison of important molecular orbitals of Keto-0 (a) and Keto-1 (b) forms with two related fragments, methyl HOBT and methyl OT. This shows the derivation of the molecular states in the Keto-0 and Keto-1 molecules. The shadowed rectangular boxes cast on Keto orbitals indicate the overlapping area between HOMO and LUMO, corresponding to the effective excitations. The higher overlap rate explains the Keto-1 molecule's high excitation intensity.

As discussed above, the location and volume of the HOMO-LUMO overlaps play an important role in the process of optical transition. A detailed analysis shows that the dominant overlapping region includes the connection between the two thiazolyl rings. In order to understand further the process of light emission, we studied OxyLH2 as if it resulted from an assembly of two fragments, HOBT (hydroxybenzothiazolyl) and OT (oxythiazolyl). The reactive sites of HOBT and OT were methylated, in substitution for the removed thiazolyl groups. Figure 2 lists the main orbitals close to HOMO and LUMO of OxyLH2, methyl HOBT and methyl OT. The orbitals from HOMO-2 to LUMO+1 in OxyLH2 can be clearly associated with orbitals in the methyl HOBT and OT functional groups. The order of some orbitals is changed, and some orbitals from different groups hybridize into one new orbital in OxyLH2. For example, the HOMO in the methyl OT appears as the HOMO-1 in Keto-0 (Figure 2a), and the LUMOs in the methyl HOBT anion and methyl OT combine into the LUMO in Keto-1 (Figure 2b). The orbitals, localized close to the ends of OxyLH2, away from the C2-C2' connecting bond, are essentially the same as those appearing in the isolated functional fragments, indicating that the basic and important features of the fragments are well preserved and brought into the whole long molecule. From the discussion above, it is clear that the most critical part of the molecule is the C2-C2' bridging bond between the two thiazolyl rings, formed by π−π conjugation. At a latter section of this work, this analysis and, in particular, the similarity between this bridging bond and a heterogeneous junction will be exploited. But, before that, we need to discuss the influence of the π−π conjugation on emitted color.

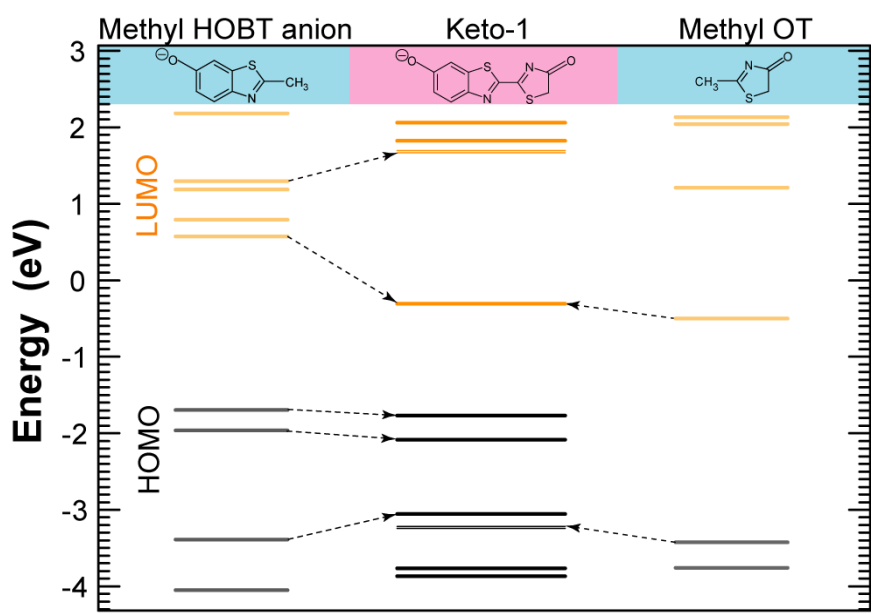

**Figure 3.** Energetic relationship of molecular states between Keto-1 and its two isolated fragments. The mapping relation of the states is shown by arrows.

Figure 3 summarizes the similarities of the Keto-1, methyl HOBT anion, and methyl OT orbitals. These states are aligned relative to a common vacuum level. One can see that those states close to the HOMO-LUMO gap are mainly coming from the HOBT group. Also, the energies of the LUMO states from the methyl HOBT anion are considerably pulled down, whereas those from the methyl OT are increased slightly. This can be explained by the formation of the π−π conjugation between two thiazolyl groups, replacing the σ−π hyperconjugation[31] between the thiazolyl and the methyl groups. As the orbital overlap in the σ−π interaction is rather weak, the π−π conjugation in Keto-1, with comparatively higher overlap, effectively stabilizes the C2-C2' bond. Thus, the higher binding energy of the C2-C2' bond (bond length 1.407 Å) in Keto-1 pulls down the LUMO of HOBT and aligns it with the LUMO of OT, thus forming the new LUMO. As a corollary of the energy downshifting, the LUMO-HOMO gap is then effectively narrowed into the energy range of visible light. In brief, the π–π conjugation between two thiazolyl groups stabilizes the OxyLH2 molecule, and brings its main peak into the visible region.

The brightness of the light emitted from a chromophore is directly associated with the intensity of the spectral peaks (Figure 1). In the firefly bioluminescence reaction, the first absorption peak (A1), which directly promotes the chromophore to the first excited state, controls the main color of emitted light. This peak occurs, for the Keto-1 and Enol-2 forms – the most plausible light emitters, in the orange－red range, in agreement with reported data.[32] From Figure 1, it is also apparent that the highest A1 signal comes from Keto-1. Its intensity is about two times higher than that of the neutral Keto-0 form and is 151% of the A1 peak intensity of the other emitting candidate, Enol-2. This indicates that Keto-1 would be the most efficient light emitter.

One cannot forget that these are gas-phase results. The environment would clearly influence the color as well as the intensity of emitted light. The analysis above is nevertheless still meaningful. In particular because, although many studies have already been devoted to the issue of color tuning, only a few authors devoted their research to the redetermination of the quantum yield, implying still large room for yield enhancement.[33] In order to develop a guideline for efficiency improvement, we propose a mechanism based on the heterojunction effect.

As mentioned before, the overlapping volume of HOMO-LUMO orbitals governs the transition probability, i.e., the spectral intensity. Another key issue discussed previously is the importance of the π–π conjugation between two thiazolyl groups. Actually, the formation of the C2-C2' bond also influences the neighboring sulfur and nitrogen atoms. The existence of the hetero-atoms, nitrogen and sulfur, breaks the homogeneity of the aromatic hydrocarbon and makes thiazole an important functional group. The heterogeneity, essentially, stems from the different electronegativities of carbon, sulfur and nitrogen, which are, respectively, 2.55, 2.58, and 3.04 in the Pauling scale.[34] In view of this, the N=C-S half ring plays a key role in the functionalization. In the case of OxyLH2, and according to the functional groups, the whole molecule can be divided in three segments, the left hand side HOBT, the right hand side OT, and the bridging part, S,N=C-C'=N',S'.

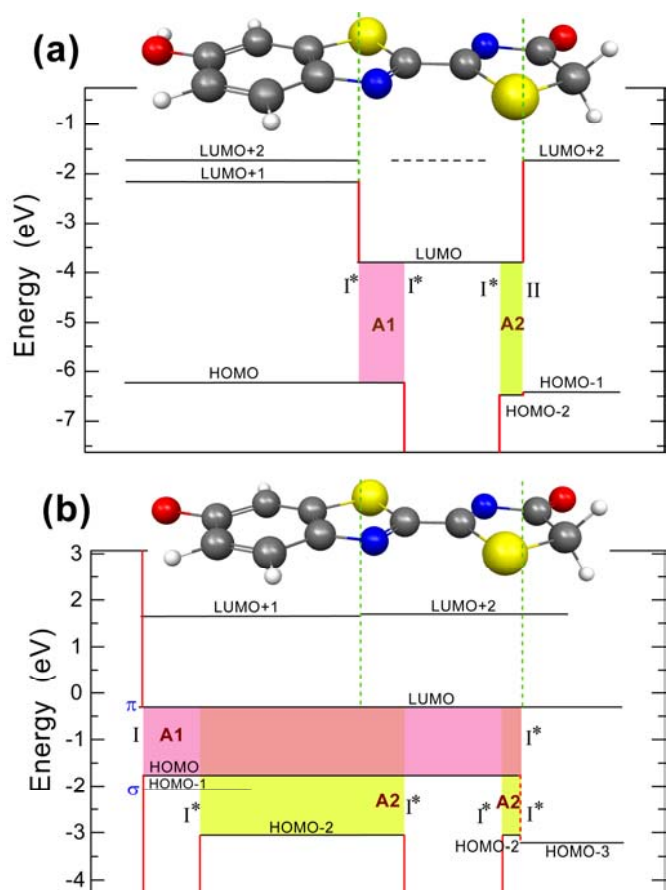

**Figure 4.** Bio-heterojunctions forming in single molecule Keto-0 (a) and Keto-1 (b). The horizontal full lines indicate the spatial distribution of the corresponding molecular orbitals along the long molecular axis. The shadowed areas show the extent of the overlap of the HOMO and LUMO orbitals chiefly responsible for the A1 and A2 transitions. The red vertical lines are located at the sites of energy barriers. Green vertical lines indicate the transition to a different functional fragment, where energy barriers usually appear. A junction forms whenever the horizontal line of HOMO or LUMO breaks and an energy offset appears between two molecular states from different sides. The type of heterojunctions I, II and I* is defined as shown in Scheme 1. The type ($\sigma$ or $\pi$) of orbitals of key molecular states is also indicated.

Based on the spatial extent of the molecule orbitals close to HOMO and LUMO, the molecular states of different segments in Keto-0 and Keto-1 are plotted in Figure 4. In this figure, the coverage of the molecular state lines show the spatial distribution of the corresponding molecular orbitals along the long molecular axis and the shadowed areas represent the transitions responsible for the corresponding optical peaks, marked A1 and A2. The junction usually forms at the coherent point between two different fragments, whenever the spatial distribution (horizontal full lines in Figure 4) of HOMO or LUMO breaks and an energy barrier appears between molecular states from two sides. One can see that the occupied and unoccupied states that contribute to the effective optical transitions are those being confined by the neighboring energy barriers, which are indicated by red vertical lines. These barriers may come from the states with higher (unoccupied) or lower (occupied) energy, or the vacuum level. Such microscopic energy barriers and wells form the atomic-level heterojunctions and introduce weak confinement on the molecular states within the overlap regions. Detailed examinations demonstrate that these bio-heterojunctions can be classified in several different types. In semiconductor physics, two representative types of heterojunctions have been defined, type I and type II, as show schematically in Scheme 1a-b,[35, 36] organized by band gap alignment. In type I heterojunctions, both occupied and unoccupied states are confined on the same side of the junction interface, whereas in type II junctions the states are confined on different sides. However, the case in molecular systems seems to be more complicated and impure. Therefore, we define another type of heterojunction, that we will refer to as type I*, as an incompletely confined type I junction (Scheme 1c-d). In principle, the strength of confinement of different heterojunctions is following the decreasing sequence from type I, I* to II. The issue of molecular junctions,[37, 38] has attracted great interest recently, with a type II heterojunction being directly observed in a single bipolar molecule by scanning tunneling microscopy.[39] One can see in Figure 4a that the states contributing to the Keto-0 A1 peak are confined between two I* junctions, while those for the A2 peak are confined on the left by a type I* and on the right by a type II junction. In Figure 4b (Keto-1) it is clear that, for the A1 peak, there is a type I/I* confinement, while the states contributing to the A2 peak are confined between two type I* junctions. As the states involved, e.g. in the π bonds, are partially delocalized over the aromatic groups, the pure type I quantum well is rarely seen in the natural molecules. However, weak confinement mostly by the type I* heterojunctions still leads to effective overlap of HOMO-LUMO states, and consequently can result in strong light emission.

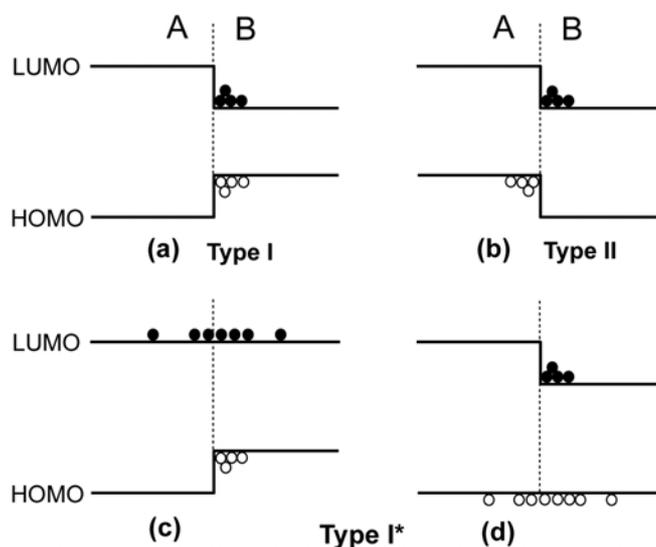

**Scheme 1.** Types of heterojunctions between fragment A and B, organized by the HOMO-LUMO gap alignment. The solid and hollow dots represent the particles confined by the heterojunctions.

The above analysis indicates that the Keto-1 form should have an increased efficiency, in accordance to the calculated optical spectra (Figure 1). Also, under the heterojunction confinement, the overlapping volume associated with the A1 transition in Keto-1 is much higher than that in Keto-0. This higher overlapping volume is another reason for the stronger spectral intensity. But we further find that the location of the overlap also plays an important role. With nitrogen being significantly more electronegative than both carbon and sulfur, the upper valence electrons will be strongly attracted to the nitrogen atoms. The charge density calculations showed that the greatest charge densities appear at the nitrogen sites. In view of this, when the overlap involves the nitrogen site, the large charge density therein would make it the most efficient optical transition center, and give rise to a sharp spectral peak. This effect has also been observed in the GaN semiconductor superlattice system,[40] strongly supporting the above conclusion. The above conclusions still apply to the Enol forms.

Finally, in order to show how to apply this proposed heterojunction concept to improve the fluorescence yield of the chromophore, an example of bioengineering modification was designed by converting the 6-hydroxy group of Keto-0 to methoxyl. The idea is to reduce the potential barrier in the left HOBT fragment through the realignment of LUMO level by inductive withdrawal of electron density, so that the heterojunction can be shifted leftwards. The results (see supporting information) showed that as expected, the LUMO level at HOBT is energetically lowered and the orbital spreads extensively from the OT ring into the HOBT ring, which increases the effective LUMO-HOMO overlap. As a corollary of the overlap improvement, the spectral intensity of A1 peak is significantly enhanced by 31%, though this enhancement is not as that much as observed in the Keto-1 form. This example indicates that the bio-heterojunction design based on molecular level re-alignment between functional groups could also be used in the applications and further development of molecular optoelectronic devices.

In summary, we computed TDDFT absorption spectra for the firefly chromophore, OxyLH2, in order to clarify the mechanism for bright light emission. The most important OxyLH2 states, the keto and enol forms, were systematically studied. Results showed that the two most intense absorption peaks are mainly associated with HOMO→LUMO and HOMO-2→LUMO transitions. We then proposed an analogy with semiconductor heterojunctions, and three types of heterojunctions, type I, type II and type I*, were introduced to account for different HOMO-LUMO gap alignments. Weak bio-heterojunction confinement on the HOMO-LUMO molecular orbitals was suggested as an explanation for the high efficiency of OxyLH2 fluorescence. Under the heterojunction confinement, the HOMO-LUMO overlapping volume turned out to be the critical criterion governing the transition probability. Due to higher overlapping volume, the efficiency of the first excitation in Keto-1 anion is significantly improved by 137 %, compared to the neutral Keto-0 form. It was also shown that the nitrogen atom in the thiazolyl ring, due to its large electronegativity, plays a key role in the chromophore fluorescence.

The bio-heterojunction confinement effect and the increase of the HOMO-LUMO overlapping volume at the nitrogen site can both be used for enhancing the fluorescence efficiency in the firefly chromophore. These concepts can also be applied to other fluorescent molecules or bipolar molecules by, e.g., functional design or side group modification, in the applications of molecular devices.

**Computational Method**

In recent years, TDDFT has emerged as a very successful framework for describing the optical spectra of nanostructured materials [41] as well as of biological systems.[22, 23] Despite its well-known shortcomings in describing systems where long-range correlation or charge transfer are important, TDDFT is nevertheless able to capture the physics of many of these processes. It is also the only electronic structure method currently available that is realistically able to tackle very large systems like the ones present in many photo-biological systems. In this work we are mainly focusing on the hetero-junction analogy and not so much on obtaining extremely accurate predictions of the optical properties of the systems being studied. TDDFT then appears as a clear choice of method, given its speed and reasonable predictive accuracy.

Gas-phase geometric optimizations of the different OxyLH2 forms were then carried out at the DFT level using the Orca package (version 2.6.35). [42] We used the B3LYP hybrid functional[43, 44] and the Kohn-Sham wavefunctions were expanded in the 6-31+G(d) basis set.[45] The resulting structures were used in all subsequent TDDFT calculations. TDDFT optical absorption spectra were calculated using the Octopus code.[46] Core electrons were kept frozen, the electron-ion interaction being described through norm-conserving pseudopotentials.[44] Exchange-correlation effects were treated within the local-density approximation (LDA),[47-49] in the Perdew-Zunger numerically stable parameterization.[50] In this OxyLH2 case, the LDA already yields good results when compared to the computationally demanding B3LYP (see supporting information). All the dynamical quantities were computed by evolving the electronic wave functions in real time and real space.[46] We used a uniform grid spacing of 0.20 Å and a time step of 0.001 fs, which ensured the stability of the time-dependent propagation and yielded spectra with a resolution better than 0.1 eV. On the other hand, to analyze the origin of the transitions responsible for the spectral maxima, we used Casida's method[51] to obtain the transition matrices. The 3D spatial extent of orbitals of optically important molecular states was also plotted. It is important to note here that the absorption spectrum (especially the first excitation) often reflects the main features of the optical properties of the inverse action (light emission) though a red shift will take place after the excited state relaxation. In fact, the emission probability can be simply described by the van Roosbroeck-Shockley relation,[23] as a function of the absorption rate $\alpha$: $R(\nu) = \rho\alpha(\nu)$, where $\rho$ is the photon density that is given by $8\pi\nu^2 n^3/c^3$.

**Acknowledgment.** The authors acknowledge support from Foundation for Science and Technology of Portugal (Project No. PTDC/FIS/73578/2006). They also thank the Centre for Computational Physics of the University of Coimbra for providing access and computer time on the milipeia supercomputer.

**Supporting Information Available**. Absorption and emission spectra with different functionals, modified heterojunctions in methyl OxyLH2 and their optical spectra. This material is available free of charge via the Internet at http://pubs.acs.org.